\documentclass[3p]{elsarticle}
\usepackage[T1]{fontenc}
\usepackage[latin9]{inputenc}
\usepackage{amsmath}
\usepackage{amssymb}
\usepackage{graphicx}
\usepackage{esint}

\makeatletter

\providecommand{\tabularnewline}{\\}

\makeatletter
\def\ps@pprintTitle{%
 \let\@oddhead\@empty
  \let\@evenhead\@empty
  \let\@oddfoot\@empty
  \let\@evenfoot\@oddfoot
}
\makeatother

\makeatother

\begin{document}
\global\long\def\dt#1{\frac{\mathrm{d}#1}{\mathrm{d}t}}

\global\long\def\Dt#1{\frac{\mathrm{D}#1}{\mathrm{D}t}}

\global\long\def\pp#1#2{\frac{\partial#1}{\partial#2}}

\global\long\def\vek#1{\boldsymbol{#1}}

\global\long\def\mm#1{\mathrm{#1}}

\title{Advection of Inertial Particles in the Presence of the History Force:\\
Higher Order Numerical Schemes}
\begin{abstract}
The equations describing the motion of finite-size particles (inertial
particles) contain in their full form the history force. This force
is represented by an integral whose accurate numerical evaluation
is rather difficult. Here, a systematic way is presented to derive
numerical integration schemes of arbitrary order for the advection
of inertial particles with the history force. This involves the numerical
evaluation of integrals with singular, but integrable, integrands.
Explicit specifications of first, second and third order schemes are
given and the accuracy and order of the schemes are verified using
known analytical solutions.\end{abstract}

\author{Anton Daitche}

\address{Institute for Theoretical Physics, Westfälische Wilhelms-Universität,
Wilhelm-Klemm-Str. 9, D-48149 Münster, Germany}

\ead{anton.d@wwu.de}

\maketitle
The advection of finite-size particles (often called inertial particles)
plays an important role in engineering \citep{crowe2011multiphase}
and in many environment-related phenomena ranging from meteorology
to oceanography, e.g. cloud microphysics \citep{Falkovich2002}. Particle-based
modeling has been applied to the formation of planetesimals in the
early solar system \citep{Bracco1999} and the aggregation and fragmentation
processes in fluid flows \citep{ZVFT2008}. Example applications are
pollutant-transport forecasting for homeland defense \citep{Benczik2002},
and the location of a toxin or biological pathogen spill (e.g. anthrax)
from outbreaks in a street canyon \citep{Tang2009}. Other recent
results indicate that inertial particles might play a role in hurricane
dynamics \citep{Haller2009} and in the feeding dynamics of certain
marine animals \citep{Haller2009b}. 

The basic equation of motion for a small spherical particle of radius
$a$ and mass $m_{p}$ in a viscous fluid is given by the Maxey-Riley
equation \citep{Maxey1983,Auton1988}:
\begin{equation}
m_{p}\dt{\vek v}=m_{f}\Dt{\vek u}-\frac{m_{f}}{2}\left(\dt{\vek v}-\Dt{\vek u}\right)-6\pi a\varrho_{f}\nu\left(\vek v-\vek u\right)-6a^{2}\varrho_{f}\sqrt{\pi\nu}\int_{t_{0}}^{t}\frac{1}{\sqrt{t-\tau}}\left(\frac{\mathrm{d}\vek v}{\mathrm{d}\tau}-\frac{\mathrm{d}\vek u}{\mathrm{d}\tau}\right)\,\mathrm{d}\tau.\label{eq:MR-Equations-dimensional}
\end{equation}
Here, $\vek v=\mathrm{d}\vek r/\mathrm{d}t$ is the particle velocity,
$\vek u(\vek r,t)$ the fluid velocity, $m_{f}$ the mass of the fluid
excluded by the particle, $\nu$ the kinematic viscosity of the fluid
and $\varrho_{f}$ the density of the fluid. The two appearing derivatives
\[
\dt{\vek u}=\pp{\vek u}t+\vek v\cdot\nabla\vek u\qquad\mbox{and}\qquad\Dt{\vek u}=\pp{\vek u}t+\vek u\cdot\nabla\vek u
\]
denote the full derivative along the trajectory of the particle and
of the corresponding fluid element, respectively. The terms on the
right-hand side of (\ref{eq:MR-Equations-dimensional}) are: the force
exerted by the fluid on a fluid element at the location of the particle,
the added mass term describing the impulsive pressure response of
the fluid, the Stokes drag, and the history force. In this form of
the equation, gravity and the so-called Faxén corrections are not
included. The history force accounts for the viscous diffusion of
vorticity from the surface of the particle along its trajectory \citep{Maxey1983}
and renders the advection equation to be an\emph{ }integro-differential
equation whose solution is much more demanding than that of an ordinary
differential equation. Because of this difficulty, this integral term
is neglected in nearly all the applications mentioned above. However,
experimental and analytic efforts \citep{Mordant2000,Angilella2004}
indicate that the history force might have significant effects for
non-neutrally-buoyant particles in simple flows. Recent studies have
also shown that the history force is relevant in turbulent flows \citep{Armenio2001,Aartrijk2010}
and chaotic advection \citep{Daitche2011}. The present paper will
detail the derivation and analysis of the numerical schemes developed
for the investigations in the latter study.

An important condition for the validity of equation (\ref{eq:MR-Equations-dimensional})
is that the particle Reynolds number $Re_{p}=\left|\vek v-\vek u\right|a/\nu$
remains small during the entire dynamics \citep{Maxey1983}. Furthermore
the particle's size $a$ and its diffusive time scale $\tau_{\nu}=a^{2}/\nu$
have to be (much) smaller then the smallest length and time scales
of the flow, respectively. For particles of comparable size as the
smallest length scale so-called Faxén corrections will become important
\citep{Maxey1983}. Several attempts \citep{Lovalenti1993,Mei1994,Dorgan2007}
have been made to extend (\ref{eq:MR-Equations-dimensional}) to the
case of finite particle Reynolds numbers by modifying the particular
form of the forces. Part of all of these approaches is a different
form of the history force. The numerical schemes presented here can
be applied to these forms as well (with some minor modifications)
as will be discussed in section \ref{sec:discussion_conclusion}.
Note also that besides the history force, further modifications of
(\ref{eq:MR-Equations-dimensional}) can be necessary for finite particle
Reynolds numbers, e.g. non-linear drag and the so-called lift force
(see \citep{Loth2009} for a review).

The history force poses the main difficulty for a numerical integration
of (\ref{eq:MR-Equations-dimensional}). There are basically three
problems: (i) the singularity of the kernel $1/\sqrt{t-\tau}$, (ii)
the fact that (\ref{eq:MR-Equations-dimensional}) is an implicit
integro-differential equation due to the appearance of $\mm d\vek v/\mm dt$
on the right-hand side and (iii) the high computational costs for
a numerical integration. The first point (i) is the most involved
one and will be addressed by a special quadrature%
\footnote{In this article the term ``quadrature scheme'' refers to a numerical
scheme for the approximation of an integral whereas the term ``integration
scheme'' refers to a scheme for the approximation of the solution
of the whole integro-differential equation.%
} scheme. The implicitness (ii) is not a major issue and can be addressed
rather easily as we will see. The last point (iii) stems from the
necessity to recompute the history force -- an integral over all previous
time-steps -- for every new time-step. Therefore the computational
costs grow with the square of the the number of time-steps and can
become quite substantial for long integration periods. This difficulty
is inherent to the dynamics governed by the history force and cannot
be addressed without further approximations. Note however that a higher
order scheme reduces the number of necessary time-steps and therefore
diminishes the problem of high computational costs indirectly. Furthermore
the final form of the numerical scheme will be formulated as a weighed
sum, which is well suited for a numerical evaluation on modern CPU
architectures.

The correct numerical treatment of the full Maxey-Riley equation and
in particular of the history force has received little interest in
the past, in spite of an increasing number of studies supporting its
importance. Michaelides~\citep{Michaelides1992} transformed the
Maxey-Riley equation to a second order equation in which the history
integral contains only the fluid velocity, but not the particle velocity.
This makes the evolution equation explicit. Furthermore, according
to Michaelides, this form of the equation allows a sparser sampling
of the particle's history, which leads to savings in computational
time and computer memory. However, the history integral still has
a similar form as in (\ref{eq:MR-Equations-dimensional}) and the
difficulties of an accurate numerical evaluation remain. Two previously
proposed schemes addressing the history integral have been tested
by Bombardelli et al. \citep{bombardelli2008}. They found the accuracy
of the schemes to be $\mathcal{O}(\sqrt{h})$ and $\mathcal{O}(h)$,
where $h$ is the time-step. In a recent work Hinsberg et al.~\citep{Hinsber2011}
have proposed a first order%
\footnote{In the paper by Hinsberg et al. the scheme is said to be of second
order. This is due to a different definition of the meaning of ``order''.
Here, a scheme with an error term proportional to the square of the
time-step is considered to be of first order as it is accurate up
to the first order; in the same sense as the Euler-method is a first
order scheme.%
} scheme for the computation of the history force, i.e. the error is
$\mathcal{O}(h^{2})$, which represents a significant advancement
over previously known schemes. Furthermore Hinsberg et al. developed
a method to decrease the needed amount of history for the computation
of the history force, by approximating the tail of the history kernel
with exponential functions. This leads to significant savings of computational
time and computer memory. This method can be viewed as a major improvement
over the method of a window kernel where the kernel is set to zero
for time lags larger then a certain window time \citep{Dorgan2007,bombardelli2008}. 

The present paper will describe the construction of arbitrary high
order methods for the integration of particle trajectories with the
history force and will give explicit specification of the first, second
and third order methods with an accuracy of $\mathcal{O}(h^{2})$,
$\mathcal{O}(h^{3})$ and $\mathcal{O}(h^{4})$, respectively. Approximate
forms of the history kernel as mentioned above will not be considered.
However, the developed schemes can be easily adapted to the window
kernel or the more advanced approach proposed by Hinsberg et al.

The rest of the paper is structured as follows: First some general
notes about the history force and the Maxey-Riley equation will given.
Afterwords a numerical quadrature scheme for the history force and
its derivation will be presented. In the following section this quadrature
scheme will be incorporated into an integration scheme for the numerical
solution of the full Maxey-Riley equation. The full integration scheme
will then be tested against known analytical solutions. This is followed
by a section on the stability properties of the algorithm, and by
a discussion and conclusion.

\section{Introductory Notes}

Measuring time and velocity in units of $T$ and $U$, the dimensionless
Maxey-Riley equation becomes

\begin{equation}
\frac{1}{R}\dt{\vek v}=\Dt{\vek u}-\frac{1}{S}\left(\vek v-\vek u\right)-\sqrt{\frac{3}{\pi}\frac{1}{S}}\int_{t_{0}}^{t}\frac{1}{\sqrt{t-\tau}}\left(\frac{\mathrm{d}\vek v}{\mathrm{d}\tau}-\frac{\mathrm{d}\vek u}{\mathrm{d}\tau}\right)\,\mathrm{d}\tau.\label{eq:MR-Equations-dimensionless}
\end{equation}
Here two dimensionless parameters appear, the density parameter%
\footnote{In some cases the density parameter is defined as $R=2m_{f}/(m_{f}+2m_{p})$,
which differs by a factor of $3/2$ from the definition here.%
} 
\[
R=\frac{3m_{f}}{m_{f}+2m_{p}},
\]
and a ratio of the particle's viscous relaxation time and the characteristic
time of the flow $T$ 
\[
S=\frac{1}{3}\frac{a^{2}/\nu}{T}.
\]
In smooth large-scale flows there is often only one typical time scale
whereas in a turbulent flow there are many. In the latter case the
smallest time scale, the Kolmogorov time $\tau_{\eta}$, is appropriate.

Many of the derivations and concepts in this article are applicable
for any kernel appearing in the history force integral. Therefore,
in the following, a general kernel $K\left(t-\tau\right)$ will be
used where the derivations do not depend on its particular form. The
explicit specification of the quadrature scheme and the tests of the
numerical schemes will be given for the standard kernel 
\begin{equation}
K(t-\tau)=\frac{1}{\sqrt{t-\tau}}.\label{eq:standard-kernel}
\end{equation}

Before we proceed with the derivation of the quadrature scheme, let
us first rewrite the history force integral in a different form
\begin{equation}
\int_{t_{0}}^{t}K\left(t-\tau\right)\frac{\mathrm{d}}{\mathrm{d}\tau}f(\tau)\,\mathrm{d}\tau+K(t-t_{0})f(t_{0})=\dt{}\int_{t_{0}}^{t}K\left(t-\tau\right)f(\tau)\,\mathrm{d}\tau,\label{eq:dt-Trick}
\end{equation}
where $f(\tau)=\vek v-\vek u$. This relation can be verified using
integration by parts%
\footnote{When the kernel has singularities, one has first to use integrals
with the upper bound of $t-\epsilon$, then perform integration by
parts and finally take the limit $\epsilon\rightarrow0$ (to prevent
the appearance of singularities outside of integrals). An alternative
for the standard kernel is to use a transformation of the integration
variable $\tau\rightarrow x=\sqrt{t-\tau}$.%
}. Equation (\ref{eq:MR-Equations-dimensional}) has been derived with
the assumption of a particle starting with the same initial velocity
as the fluid, i.e. $\vek v(t_{0})=\vek u(t_{0})$. In this case the
second term on the left-hand side of (\ref{eq:dt-Trick}) vanishes.
In the case of different initial velocity the additional term $\left(\vek u(t_{0})-\vek v(t_{0})\right)/\sqrt{t-t_{0}}$
has been given in \citep{Michaelides1992,Maxey1993}, which is exactly
the additional term appearing in (\ref{eq:dt-Trick}). Therefore the
Maxey-Riley equation can be written in the following form, which is
now also valid for initial conditions with $\vek v(t_{0})\neq\vek u(t_{0})$,
\begin{equation}
\frac{1}{R}\dt{\vek v}=\Dt{\vek u}-\frac{1}{S}\left(\vek v-\vek u\right)-\sqrt{\frac{3}{\pi}\frac{1}{S}}\,\dt{}\int_{t_{0}}^{t}\mathrm{d}\tau\, K(t-\tau)\left(\vek v-\vek u\right).\label{eq:MR-Equations-dimensionless-2}
\end{equation}
It is beneficial to use this form of the history force because it
enables us to compute an integral of the history force by simply dropping
the derivative. This improves and simplifies the numerical scheme
as we will see. 

At this point it is interesting to note that for the standard kernel
the history force is equal to a fractional derivative of the Riemann-Liouville
type: 
\[
\left(\dt{}\right)^{1/2}f(t)\equiv\frac{1}{\sqrt{\pi}}\dt{}\int_{t_{0}}^{t}\frac{1}{\sqrt{t-\tau}}f(\tau)\,\mathrm{d}\tau.
\]
Thus the numerical methods developed here can be also considered as
higher order methods for the numerical computation of fractional derivatives
and the solution of fractional differential equations.

\section{The Quadrature Scheme\label{sec:Quadrature-Scheme}}

In this section a systematic way is presented for the construction
of quadrature schemes for integrals of the type 
\[
\int_{t_{0}}^{t}K\left(t-\tau\right)f(\tau)\,\mathrm{d}\tau.
\]
When the kernel $K$ is a well behaved function no special effort
is needed and standard schemes can be used. However in cases where
the kernel has an integrable singularity, like e.g. the standard kernel
(\ref{eq:standard-kernel}) at $\tau=t$, standard numerical methods,
like the Newton-Cotes%
\footnote{Well known Newton-Cotes schemes are e.g. the trapezoidal rule and
Simpson's rule. %
} schemes, lead to large errors as we will see. This is due to the
necessity to evaluate the whole integrand including the kernel near
the singularity. We will avoid this by constructing a specialized
scheme in which the kernel is already integrated analytically.

Due to the linearity of the history integral with respect to $f$
any quadrature scheme for this term can be expressed as a weighted
sum 
\[
\int_{t_{0}}^{t}K\left(t-\tau\right)f(\tau)\,\mathrm{d}\tau\approx\sum_{j=0}^{n}\mu_{j}f(\tau_{j}),
\]
where $\tau_{i}=t_{0}+hi$, $n=\left(t-t_{0}\right)/h$ and $h$ is
the time-step. The main topic of this section is the derivation and
specification of the coefficients $\mu_{j}$. The general procedure
is to first split the integral into intervals of length $h$
\[
\int_{t_{0}}^{t}K\left(t-\tau\right)f(\tau)\,\mathrm{d}\tau=\sum_{i=0}^{n-1}\int_{\tau_{i}}^{\tau_{i+1}}K\left(t-\tau\right)f(\tau)\,\mathrm{d}\tau,
\]
then to approximate $f(\tau)$ in every of the intervals with a polynomial
and finally to compute the appearing integrals analytically. The order
of the polynomial will determine the order of the scheme. 

Let us first examine the simplest case: a linear approximation leading
to an order one scheme. By approximating%
\footnote{The error of an approximation will be denoted by $\mathcal{O}(h^{m})$,
i.e. the error is bounded by $Ch^{m}$ for some fixed $C$. %
} $f(\tau)$ linearly in the interval $\left[\tau_{i},\tau_{i+1}\right]$
\begin{equation}
f(\tau)=f(\tau_{i})+\frac{f(\tau_{i+1})-f(\tau_{i})}{h}\left(\tau-\tau_{i}\right)+\mathcal{O}(h^{2})\label{eq:first-order-polynomial-interpolation}
\end{equation}
we obtain 
\[
\int_{\tau_{i}}^{\tau_{i+1}}K\left(t-\tau\right)f(\tau)\,\mathrm{d}\tau=\left(f(\tau_{i})+\mathcal{O}(h^{2})\right)\int_{0}^{h}K\left(t-\tau_{i}-\tau\right)\,\mathrm{d}\tau+\frac{f(\tau_{i+1})-f(\tau_{i})}{h}\int_{0}^{h}\tau K\left(t-\tau_{i}-\tau\right)\,\mathrm{d}\tau.
\]
In many cases the appearing integrals can be computed analytically,
e.g. for the standard kernel (\ref{eq:standard-kernel}) 
\[
\int_{\tau_{i}}^{\tau_{i+1}}\frac{f(\tau)}{\sqrt{t-\tau}}\,\mathrm{d}\tau=\left(f(\tau_{i})+\mathcal{O}(h^{2})\right)\left[-2\sqrt{t-\tau_{i}-\tau}\right]_{0}^{h}+\frac{f(\tau_{i+1})-f(\tau_{i})}{h}\left[-2\tau\sqrt{t-\tau_{i}-\tau}-\frac{4}{3}\left(t-\tau_{i}-\tau\right)^{\frac{3}{2}}\right]_{0}^{h}.
\]
Summing up the terms for each of the intervals we obtain a formula
for the whole integral, e.g. for the standard kernel
\begin{multline}
\int_{t_{0}}^{t}\frac{f(\tau)}{\sqrt{t-\tau}}\mathrm{d}\tau=2f(t_{0})\sqrt{t-t_{0}}+\frac{4}{3}\sum_{i=0}^{n-1}\frac{f(\tau_{i+1})-f(\tau_{i})}{h}\left(\left(t-\tau_{i}\right)^{\frac{3}{2}}-\left(t-\tau_{i+1}\right)^{\frac{3}{2}}\right)+\mathcal{O}(h^{2})\sqrt{t-t_{0}}.\label{eq:qaudrature-first-order}
\end{multline}
Note that no singular or diverging expressions appear. For this it
is crucial to approximate only $f(\tau)$ with polynomials, but not
the whole integrand. 

As already mentioned the quadrature scheme is linear in $f$ and can
thus be expressed as a weighted sum. Such a form is best suited for
a numerical evaluation as modern processors/compilers can optimize
this kind of operations rather well. We will index the coefficients
of the sum in reversed order, i.e. we use the sum $\sum_{j}\mu{}_{j}f(\tau_{n-j})$
instead of $\sum_{j}\mu{}_{j}f(\tau_{j})$. This is more natural as
it turns out that the coefficient of $f(\tau_{j})$ depends on $n-j$.
For the standard kernel the coefficients for the first order quadrature
scheme can be obtained from (\ref{eq:qaudrature-first-order}):

\begin{align}
\int_{t_{0}}^{t}\frac{f(\tau)}{\sqrt{t-\tau}}\mathrm{d}\tau & =\sqrt{h}\sum_{j=0}^{n}\alpha_{j}^{n}f(\tau_{n-j})+\mathcal{O}(h^{2})\sqrt{t-t_{0}}\label{eq:first-order-quadrature}\\
\alpha_{j}^{n} & =\frac{4}{3}\begin{cases}
1 & j=0\\
\left(j-1\right)^{3/2}+\left(j+1\right)^{3/2}-2j^{3/2} & 0<j<n\\
\left(n-1\right)^{3/2}-n^{3/2}+\frac{6}{4}\sqrt{n} & j=n.
\end{cases}\label{eq:first-order-coefficients}
\end{align}
Here the factor $\sqrt{h}$ has been pulled out of the coefficients
to make them independent of the time-step $h$. Also, note that the
coefficients $\alpha_{j}^{n}$ depend on $n$, the number of intervals
for the approximation of the integral. The first order scheme specified
by (\ref{eq:first-order-quadrature}) and (\ref{eq:first-order-coefficients})
is equivalent to the one presented in \citep{Hinsber2011}, although
the equivalence is not obvious.

The procedure just shown can be generalized to derive quadrature schemes
of arbitrary high order. The basic ideas stay the same, however the
technical details make the derivation complicated. Here, only a simplified
overview of the construction will be given. The full derivation with
all the technical details can be found in \ref{sec:appendix-1}.

To obtain a quadrature scheme of order $m$, we approximate $f$ in
every interval $\left[\tau_{i},\tau_{i+1}\right]$ with an $m$\nobreakdash-th
order polynomial and solve the remaining integrals analytically. An
interpolating polynomial of order $m$ is uniquely determined by the
values of $f$ at $m+1$ time-points. Let us denote these time-points
by $\theta_{ik}$, where $i$ is the index of the interval and $k\in\left\{ 0,\ldots,m\right\} $.
Using the Lagrangian representation of polynomial interpolation we
obtain the approximation in the $i$-th interval 
\[
f(\tau)=\sum_{k=0}^{m}f(\theta_{ik})L_{ik}\left(\tau\right)+\mathcal{O}(h^{m+1})\qquad L_{ik}(\tau)=\prod_{\substack{l=0\\
l\neq k
}
}^{m}\frac{\tau-\theta_{il}}{\theta_{ik}-\theta_{il}}.
\]
The time-points $\theta_{ik}$ can in principle be chosen arbitrary.
However, it is clear that this choice will strongly influence the
quality of the interpolation. Obviously, the points $\tau_{i}$ and
$\tau_{i+1}$ should be included when interpolating in $\left[\tau_{i},\tau_{i+1}\right]$.
These time-points were our choice for the first order approximation
(\ref{eq:first-order-polynomial-interpolation}). For higher order
approximations we need more points additionally to $\tau_{i}$ and
$\tau_{i+1}$. Reasonable choices are the points closest to the bounds
of the interval, i.e. $\tau_{i-1}$, $\tau_{i+2}$, $\tau_{i-2}$,
$\ldots$ (given we want to stay on the grid defined by the $\tau_{i}$).
And indeed we will use $\tau_{i-1}$, $\tau_{i}$, $\tau_{i+1}$ for
the second order approximation and $\tau_{i-1}$, $\tau_{i}$, $\tau_{i+1}$,
$\tau_{i+2}$ for the third order approximation. This can be generalized
to arbitrary orders by choosing $\theta_{ik}=\tau_{i-\left\lfloor m/2\right\rfloor +k}$
where the operation $\left\lfloor \cdot\right\rfloor $ denotes taking
the integer part, often called the floor function.

With this definitions we can express the history integral as
\[
\int_{t_{0}}^{t}\mathrm{d}\tau\, K\left(t-\tau\right)f(\tau)=\sum_{i=0}^{n-1}\sum_{k=0}^{m}f(\theta_{ik})\underbrace{\int_{\tau_{i}}^{\tau_{i+1}}\mathrm{d}\tau\, K\left(t-\tau\right)L_{k}(\tau)}_{\lambda_{ik}}+E=\sum_{i=0}^{n-1}\sum_{k=0}^{m}f(\theta_{ik})\lambda_{ik}+E,
\]
where $E=\int_{t_{0}}^{t}K(t-\tau){\rm d}\tau\,\mathcal{O}(h^{m+1})$
is the error term. We naturally obtain a weighted double-sum due to
the use of the Lagrangian representation of the interpolating polynomial,
where the $f(\theta_{ik})$ appear as coefficients of the polynomials
$L_{ik}$. Compare this with derivation of the first order scheme
where we started from the linear interpolation (\ref{eq:first-order-polynomial-interpolation}),
which is not in the Lagrangian form and thus a reordering of terms
was necessary to get from (\ref{eq:qaudrature-first-order}) to (\ref{eq:first-order-coefficients}). 

The integrals $\lambda_{ik}$ do not involve $f(\tau)$ and can be
computed in advance; for many kernels even analytically, including
the standard kernel. Now the final step is to reorder the double sum
to a single weighted sum
\[
\int_{t_{0}}^{t}\mathrm{d}\tau\, K\left(t-\tau\right)f(\tau)=\sum_{i=0}^{n-1}\sum_{k=0}^{m}f(\theta_{ik})\lambda_{ik}=\sum_{j=0}^{n}\mu_{j}^{n}f(\tau_{n-j}).
\]
This procedure is detailed in \ref{sec:appendix-1}. Note that the
coefficients $\mu_{j}^{n}$ (just like $\alpha_{j}^{n}$) have a dependence
on $n$.

In the following the coefficients for the standard kernel (\ref{eq:standard-kernel})
will be given for a second and third order schemes, denoted by $\beta_{j}^{n}$
and $\gamma_{j}^{n}$ respectively. The factor $\sqrt{h}$ has been
extracted from the coefficients so that they do not depend on the
time-step $h$.

The second order approximation is 
\begin{equation}
\int_{t_{0}}^{t}\mathrm{d}\tau\,\frac{1}{\sqrt{t-\tau}}f(\tau)=\sqrt{h}\sum_{j=0}^{n}\beta_{j}^{n}f(\tau_{n-j})+\mathcal{O}(h^{3})\sqrt{t-t_{0}}\label{eq:second-order-quadrature}
\end{equation}
with $\beta_{j}^{n}$ for $n=2$ and $n=3$
\begin{eqnarray*}
\beta_{j=0,1,2}^{2} & = & \frac{12}{15}\sqrt{2};\;\frac{16}{15}\sqrt{2};\;\frac{2}{15}\sqrt{2}\\
\beta_{j=0,1,2,3}^{3} & = & \frac{4}{5}\sqrt{2};\;\frac{14}{5}\sqrt{3}-\frac{12}{5}\sqrt{2};\;-\frac{8}{5}\sqrt{3}+\frac{12}{5}\sqrt{2};\;\frac{4}{5}\sqrt{3}-\frac{4}{5}\sqrt{2}
\end{eqnarray*}
and for $n\geq4$

\[
\beta_{j}^{n}=\begin{cases}
\frac{4}{5}\sqrt{2} & j=0\\
\frac{14}{5}\sqrt{3}-\frac{12}{5}\sqrt{2} & j=1\\
\frac{176}{15}-\frac{42}{5}\sqrt{3}+\frac{12}{5}\sqrt{2} & j=2\\
\\
\frac{8}{15}\left(\left(j+2\right)^{5/2}-3\left(j+1\right)^{5/2}+3j^{5/2}-\left(j-1\right)^{5/2}\right) & 2<j<n-1\\
+\frac{2}{3}\left(-\left(j+2\right)^{3/2}+3\left(j+1\right)^{3/2}-3j^{3/2}+\left(j-1\right)^{3/2}\right)\\
\\
\frac{8}{15}\left(-2n^{5/2}+3\left(n-1\right)^{5/2}-\left(n-2\right)^{5/2}\right) & j=n-1\\
+\frac{2}{3}\left(4n^{3/2}-3\left(n-1\right)^{3/2}+\left(n-2\right)^{3/2}\right)\\
\\
\frac{8}{15}\left(n^{5/2}-\left(n-1\right)^{5/2}\right)+\frac{2}{3}\left(-3n^{3/2}+\left(n-1\right)^{3/2}\right)+2\sqrt{n} & j=n.
\end{cases}
\]

The third order approximation is
\begin{equation}
\int_{t_{0}}^{t}\mathrm{d}\tau\,\frac{1}{\sqrt{t-\tau}}f(\tau)=\sqrt{h}\sum_{j=0}^{n}\gamma_{j}^{n}f(\tau_{n-j})+\mathcal{O}(h^{4})\sqrt{t-t_{0}}\label{eq:third-order-quadrature}
\end{equation}
with $\gamma_{j}^{n}$ for $3\leq n\leq6${\small 
\begin{eqnarray*}
\gamma_{j=0..3}^{3} & = & \frac{68}{105}\sqrt{3};\:\frac{6}{7}\sqrt{3};\:\frac{12}{35}\sqrt{3};\:\frac{16}{105}\sqrt{3}\\
\gamma_{j=0..4}^{4} & = & \frac{244}{315}\sqrt{2};\:\frac{1888}{315}-\frac{976}{315}\:\sqrt{2};\:-\frac{656}{105}+\frac{488}{105}\sqrt{2};\:\frac{544}{105}-\frac{976}{315}\sqrt{2};\:-\frac{292}{315}+\frac{244}{315}\:\sqrt{2}\\
\gamma_{j=0..5}^{5} & = & \frac{244}{315}\sqrt{2};\:\frac{362}{105}\,\sqrt{3}-\frac{976}{315}\sqrt{2};\:\frac{500}{63}\sqrt{5}-\frac{1448}{105}\sqrt{3}+\frac{488}{105}\sqrt{2};\:-\frac{290}{21}\sqrt{5}+\frac{724}{35}\sqrt{3}-\frac{976}{315}\sqrt{2};\\
 &  & \frac{220}{21}\sqrt{5}-\frac{1448}{105}\sqrt{3}+\frac{244}{315}\sqrt{2};\:-\frac{164}{63}\sqrt{5}+\frac{362}{105}\sqrt{3}\\
\gamma_{j=0..6}^{6} & = & \frac{244}{315}\sqrt{2};\:\frac{362}{105}\sqrt{3}-\frac{976}{315}\sqrt{2};\:\frac{5584}{315}-\frac{1448}{105}\,\sqrt{3}+\frac{488}{105}\sqrt{2};\:\frac{344}{21}\sqrt{6}-\frac{22336}{315}+\frac{724}{35}\sqrt{3}-\frac{976}{315}\sqrt{2};\\
 &  & -\frac{1188}{35}\sqrt{6}+\frac{11168}{105}-\frac{1448}{105}\sqrt{3}+\frac{244}{315}\sqrt{2};\:\frac{936}{35}\sqrt{6}-\frac{22336}{315}+\frac{362}{105}\sqrt{3};-\frac{754}{105}\sqrt{6}+\frac{5584}{315}
\end{eqnarray*}
}and for $n\geq7$

{\small 
\[
\gamma_{j}^{n}=\begin{cases}
\frac{244}{315}\sqrt{2} & j=0\\
\frac{362}{105}\sqrt{3}-\frac{976}{315}\sqrt{2} & j=1\\
\frac{5584}{315}-\frac{1448}{105}\sqrt{3}+\frac{488}{105}\sqrt{2} & j=2\\
\frac{1130}{63}\sqrt{5}-\frac{22336}{315}+\frac{724}{35}\sqrt{3}-\frac{976}{315}\sqrt{2} & j=3\\
\\
\frac{16}{105}\left(\left(j+2\right)^{7/2}+\left(j-2\right)^{7/2}-4\left(j+1\right)^{7/2}-4\left(j-1\right)^{7/2}+6j^{7/2}\right) & 3<j<n-3\\
+\frac{2}{9}\left(4\left(j+1\right)^{3/2}+4\left(j-1\right)^{3/2}-\left(j+2\right)^{3/2}-\left(j-2\right)^{3/2}-6j^{3/2}\right)\\
\\
\frac{16}{105}\left(n^{7/2}-4\left(n-2\right)^{7/2}+6\left(n-3\right)^{7/2}-4\left(n-4\right)^{7/2}+\left(n-5\right)^{7/2}\right)-\frac{8}{15}n^{5/2} & j=n-3\\
+\frac{4}{9}n^{3/2}+\frac{8}{9}\left(n-2\right)^{3/2}-\frac{4}{3}\left(n-3\right)^{3/2}+\frac{8}{9}\left(n-4\right)^{3/2}-\frac{2}{9}\left(n-5\right)^{3/2}\\
\\
\frac{16}{105}\left(\left(n-4\right)^{7/2}-4\left(n-3\right)^{7/2}+6\left(n-2\right)^{7/2}-3n^{7/2}\right)+\frac{32}{15}n^{5/2} & j=n-2\\
-2n^{3/2}-\frac{4}{3}\left(n-2\right)^{3/2}+\frac{8}{9}\left(n-3\right)^{3/2}-\frac{2}{9}\left(n-4\right)^{3/2}\\
\\
\frac{16}{105}\left(3n^{7/2}-4\left(n-2\right)^{7/2}+\left(n-3\right)^{7/2}\right)-\frac{8}{3}n^{5/2}+4n^{3/2}+\frac{8}{9}\left(n-2\right)^{3/2}-\frac{2}{9}\left(n-3\right)^{3/2} & j=n-1\\
\frac{16}{105}\left(\left(n-2\right)^{7/2}-n^{7/2}\right)+\frac{16}{15}n^{5/2}-\frac{22}{9}n^{3/2}-\frac{2}{9}\left(n-2\right)^{3/2}+2\sqrt{n} & j=n.
\end{cases}
\]
}{\small \par}

\begin{figure}
\begin{centering}
\includegraphics[height=0.3\paperwidth]{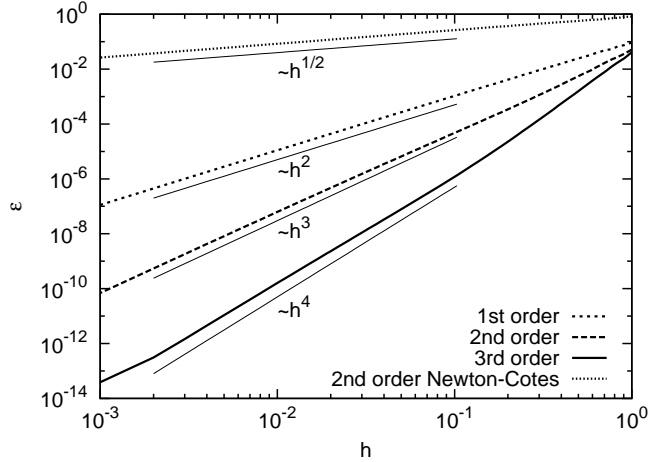}
\par\end{centering}

\caption{\label{fig:error-scaling}Scaling of the global error $\varepsilon(h)$
of the quadrature schemes for the test-case $f(\tau)=\sin(\tau)$.}

\end{figure}

Now that the quadrature schemes are fully specified, let us verify
the correctness of the derivation and in particular the order of the
schemes by using a test-case where the analytical solution of the
integral is known. We choose the case $f(\tau)=\sin(\tau)$ where
the history integral can be computed with the help of the Anger function
$J_{\nu}(t)$ \citep{Abramowitz1972}, which is a generalization of
the Bessel function $J_{n}(t)$ to fractional values of $n$,
\begin{equation}
\int_{0}^{t}\frac{\sin(\tau)}{\sqrt{t-\tau}}\mathrm{d}\tau=\frac{1}{2}\pi\sqrt{t}\left(J_{\frac{1}{2}}\left(t\right)-J_{-\frac{1}{2}}\left(t\right)\right)\equiv I(t).\label{eq:sin-exact-integral}
\end{equation}
To verify the order of the scheme let us analyze the global error
\[
\varepsilon(h)=\underset{t\in[0,10]}{\max}\left|I(t)-I_{\mathrm{num}}(t,h)\right|,
\]
where $I(t)$ denotes the exact value of the integral in (\ref{eq:sin-exact-integral})
and $I_{\mathrm{num}}(t,h)$ the numerically approximated value. Fig.~\ref{fig:error-scaling}
shows the dependence of this global error on $h$ for the three numerical
quadrature schemes given here (specified by $\alpha_{j}^{n}$, $\beta_{j}^{n}$,
$\gamma_{j}^{n}$) and a second order, semi-open Newton-Cotes scheme
\citep{Press1987}. We see that errors of the schemes are proportional
to $h^{m+1}$ for the $m$-th order scheme, thus verifying the order
of the quadrature schemes (at least for this test-case). Also we see
that a standard second order quadrature scheme (the Newton-Cotes scheme)
performs very badly as the error scales only with $\sqrt{h}$. This
is also true for higher order Newton-Cotes schemes and is due to the
necessity of a numerical evaluation of the kernel near the singularity. 

The correctness of the quadrature schemes has also been tested using
the analytically treatable case of a polynomial of arbitrary order
and led to similar results.

\section{Integration of the full Maxey-Riley equation\label{sec:inegrator-section}}

In this section the quadrature scheme developed in the previous section
will be incorporated in a multi-step integration scheme for the full
Maxey-Riley equation. To this end we formulate the Maxey-Riley equation
for the velocity difference $\vek w=\vek v-\vek u$ in a given flow
field $\vek u$:
\begin{equation}
\dt{\vek w}=\left(R-1\right)\dt{\vek u}-R\vek w\cdot\nabla\vek u-\frac{R}{S}\vek w-R\sqrt{\frac{3}{\pi S}}\dt{}\int_{t_{0}}^{t}K(t-\tau)\vek w(\tau)\,\mm d\tau.\label{eq:MR-for-w}
\end{equation}
Together with the evolution equation for the particle position 
\[
\dt{\vek r}=\vek v=\vek w+\vek u
\]
 equation (\ref{eq:MR-for-w}) fully specifies the motion of an inertial
particle in a fluid. Integrating (\ref{eq:MR-for-w}) from $t$ to
$t+h$ and using the abbreviations 
\begin{align*}
\vek G & =\left(R-1\right)\dt{\vek u}-R\vek w\cdot\nabla\vek u-\frac{R}{S}\vek w\\
\vek H & =-R\sqrt{\frac{3}{\pi S}}\int_{t_{0}}^{t}K(t-\tau)\vek w(\tau)\,\mm d\tau
\end{align*}
we obtain
\begin{equation}
\vek w(t+h)=\vek w(t)+\int_{t}^{t+h}\vek G(\tau)\,{\rm d}\tau+\vek H(t+h)-\vek H(t).\label{eq:MR-integrated}
\end{equation}
Here the integration of the history term can be performed trivially
due to relation (\ref{eq:dt-Trick}). This simplifies the integration
scheme considerably. Furthermore, we now have to compute a history
integral of $\vek w$ and not $\mm d\vek w/\mm d\tau$, where the
former will generally fluctuate less and is therefore better suited
for a numerical quadrature. The history integral $\vek H$ can be
computed with the schemes developed in section~\ref{sec:Quadrature-Scheme}.
The integral of $\vek G$ can be approximated using polynomial interpolation.
We use only the present and previous values of $\vek G$ for this
approximation in order to obtain an explicit scheme:
\begin{align*}
\int_{t}^{t+h}\vek G(\tau)\,{\rm d}\tau & =h\vek G(t)+\mathcal{O}(h^{2})\\
\int_{t}^{t+h}\vek G(\tau)\,{\rm d}\tau & =\frac{h}{2}\left(3\vek G(t)-\vek G(t-h)\right)+\mathcal{O}(h^{3})\\
\int_{t}^{t+h}\vek G(\tau)\,{\rm d}\tau & =\frac{h}{12}\left(23\vek G(t)-16\vek G(t-h)+5\vek G(t-2h)\right)+\mathcal{O}(h^{4}).
\end{align*}
These expressions can be found in any literature on Adams-Bashforth
multi-step methods, e.g. \citep{Press1987}.

A final point which we have to consider before writing down the complete
scheme, is that 
\[
\vek H(t+h)=\sum_{j=0}^{n+1}\mu_{j}^{n+1}\vek w(\tau_{n+1-j})+\mathcal{O}(h^{m})
\]
depends on $\vek w(\tau_{n+1})=\vek w(t+h)$ and thus can not be evaluated
before $\vek w(t+h)$ is known. This is due to the implicitness of
the Maxey-Riley equation. However this is easily dealt with by bringing
$\vek w(t+h)$ to the left-hand side of (\ref{eq:MR-integrated}).

If we now consider (\ref{eq:MR-integrated}) on the grid $t_{n}=t_{0}+nh$,
define $\xi=R\sqrt{3/(\pi S)}\sqrt{h}$ and use abbreviations of the
type $\vek w_{n}=\vek w(t_{n})$ we can specify the complete integration
schemes of first, second and third order for the Maxey-Riley equation:
\begin{align}
\vek r_{n+1} & =\vek r_{n}+h\left(\vek w_{n}+\vek u_{n}\right)+\mathcal{O}(h^{2}),\nonumber \\
\left(1+\xi\alpha_{0}^{n+1}\right)\,\vek w_{n+1} & =\vek w_{n}+h\vek G_{n}-\xi\sum_{j=0}^{n}\left(\alpha_{j+1}^{n+1}\vek w_{n-j}-\alpha_{j}^{n}\vek w_{n-j}\right)+\sqrt{t_{n}-t_{0}}\mathcal{O}(h^{2}),\label{eq:first-order-integrator}\\
\nonumber \\
\vek r_{n+1} & =\vek r_{n}+\frac{h}{2}\left(3\left(\vek w_{n}+\vek u_{n}\right)-\left(\vek w_{n-1}+\vek u_{n-1}\right)\right)+\mathcal{O}(h^{3}),\nonumber \\
\left(1+\xi\beta_{0}^{n+1}\right)\,\vek w_{n+1} & =\vek w_{n}+\frac{h}{2}\left(3\vek G_{n}-\vek G_{n-1}\right)-\xi\sum_{j=0}^{n}\left(\beta_{j+1}^{n+1}\vek w_{n-j}-\beta_{j}^{n}\vek w_{n-j}\right)+\sqrt{t_{n}-t_{0}}\mathcal{O}(h^{3}),\label{eq:second-order-integrator}\\
\nonumber \\
\vek r_{n+1} & =\vek r_{n}+\frac{h}{12}\left(23\left(\vek w_{n}+\vek u_{n}\right)-16\left(\vek w_{n-1}+\vek u_{n-1}\right)+5\left(\vek w_{n-2}+\vek u_{n-2}\right)\right)+\mathcal{O}(h^{4}),\nonumber \\
\left(1+\xi\gamma_{0}^{n+1}\right)\,\vek w_{n+1} & =\vek w_{n}+\frac{h}{12}\left(23\vek G_{n}-16\vek G_{n-1}+5\vek G_{n-2}\right)-\xi\sum_{j=0}^{n}\left(\gamma_{j+1}^{n+1}\vek w_{n-j}-\gamma_{j}^{n}\vek w_{n-j}\right)+\sqrt{t_{n}-t_{0}}\mathcal{O}(h^{4}).\label{eq:third-order-integrator}
\end{align}

The coefficients $\alpha_{j}^{n}$, $\beta_{j}^{n}$ and $\gamma_{j}^{n}$
are given in section~\ref{sec:Quadrature-Scheme}. One reason to
include the first and second order schemes here (besides the third
order one) is that one cannot start the integration with the third
order scheme as the previous values $\vek G_{n-1}$, $\vek G_{n-2}$
are not available at the beginning. This is a problem common to all
multi-step methods. The simplest solution is to use the first and
second order schemes for the first two steps and the third order scheme
for the rest. To perform the first step of the integration ($n=0$)
with (\ref{eq:first-order-integrator}) the coefficients $\alpha_{j}^{0}$
are needed, which we define to be zero as no history is present at
$t=t_{0}$. Ideally, we would perform the second step ($n=1$) with
(\ref{eq:second-order-integrator}). But $\beta_{j}^{n}$ is defined
only for $n\geq2$, leaving us with two options: (i) perform the second
step with the first order scheme or (ii) define $\beta_{j}^{1}\equiv\alpha_{j}^{1}$
and accept a reduced accuracy. The second option is at least as accurate
as the first one and will thus be assumes in the following. The same
considerations are applicable to the third order scheme (\ref{eq:third-order-integrator}),
leading to the definition $\gamma_{j}^{2}\equiv\beta_{j}^{2}$ and
allowing its use for $n\geq2$ (instead of for $n\geq3$).

The integration methods (\ref{eq:first-order-integrator})-(\ref{eq:third-order-integrator})
can be viewed as an extension of the Adams-Bashforth multi-step methods
to the case of an integro-differential equation with memory. In its
present form the quadrature schemes in section \ref{sec:Quadrature-Scheme}
are best suited for multi-step methods with a fixed time-step. In
other schemes, for example Runge-Kutta, half-steps are necessary,
but they cannot be evaluated with the current formulation of the quadrature
schemes. Furthermore, multi-step methods allow to profit from the
fact that an integral of the history force can be computed by simply
dropping a derivative (see (\ref{eq:MR-integrated}) and comments
below).

\subsection{Comments on the Implementation }

\begin{figure}
\begin{centering}
\includegraphics[width=0.3\paperwidth]{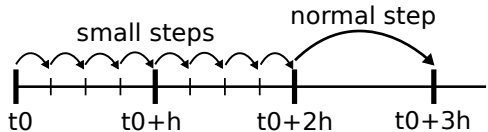}
\par\end{centering}

\caption{\label{fig:integration-start}Procedure to start the integration with
multi-step methods.}
\end{figure}

Using lower order schemes for the first two steps makes them less
accurate. A more advanced strategy is to begin the integration with
a smaller time-step to account for the reduced accuracy and switch
on the third order scheme with the normal time-step when it is applicable.
This procedure is demonstrated in figure \ref{fig:integration-start}:
At the beginning of the integration, eight small steps with time-step
$h'=h/4$ are taken. From the time-point $t_{0}+2h$ on the third
order scheme can be applied with the normal step size $h$ as enough
previous values are available then. In practice the size $h'$ of
the small steps can be much smaller than $h$, e.g. $h'=h/100$. This
procedure can be further refined, e.g. in figure \ref{fig:integration-start}
it would be sufficient to take steps of $h/2$ in the interval $\left[t_{0}+h,t_{0}+2h\right]$.
However, the savings in computational time due to this optimization
will generally be not large enough to compensate for the increased
complexity of the algorithm.

Because of the dependency of the coefficients $\alpha_{j}^{n}$, $\beta_{j}^{n}$
and $\gamma_{j}^{n}$ on $n$ one might be tempted to recompute them
for every time-step $n$. This would make the schemes quite slow.
Fortunately the coefficients can be precomputed and stored efficiently,
because they depend on $n$ only for the last few $j$. We will exemplify
this for $\alpha_{j}^{n}$ given in (\ref{eq:first-order-coefficients});
the generalization to the higher order coefficients $\beta_{j}^{n}$
and $\gamma_{j}^{n}$ is straightforward. As $\alpha_{j}^{n}$ depends
on $n$ only when $j=n$, we can express it as 
\[
\alpha_{j}^{n}=\begin{cases}
a_{j} & j<n\\
b_{n} & j=n
\end{cases}\qquad\mbox{with}\qquad a_{j}=\frac{4}{3}\begin{cases}
1 & j=0\\
(j-1)^{3/2}+(j+1)^{3/2}-2j^{3/2} & j>0
\end{cases}
\]
and $b_{n}=\frac{4}{3}\left(\left(n-1\right)^{3/2}-n^{3/2}+\frac{6}{4}\sqrt{n}\right)$.
Let now $N$ be the maximal number of time-steps we wish to perform.
We then can precompute $a_{j}$ and $b_{n}$ for $j,n\leq N$ once
and easily construct $\alpha_{j}^{n}$ from them for every $n\leq N$.
This is particularly beneficial when one wants to integrate a large
number of particle trajectories. As the coefficients $\alpha_{j}^{n}$,
$\beta_{j}^{n}$ and $\gamma_{j}^{n}$ contain differences of large
numbers (for large $j$), they should be precomputed with a high numerical
precision. For the examples shown here they have been computed with
quad precision (i.e. 128-bit floating point number) and stored with
double precision.

\subsection{Testing the Accuracy of the Schemes}

\begin{figure}
\begin{centering}
\includegraphics[height=0.3\paperwidth]{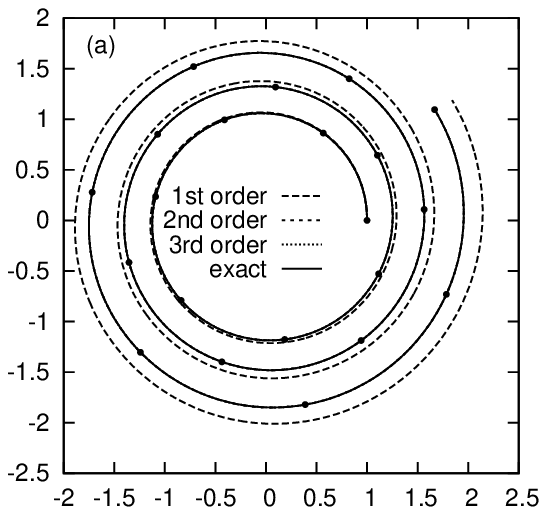}\includegraphics[height=0.3\paperwidth]{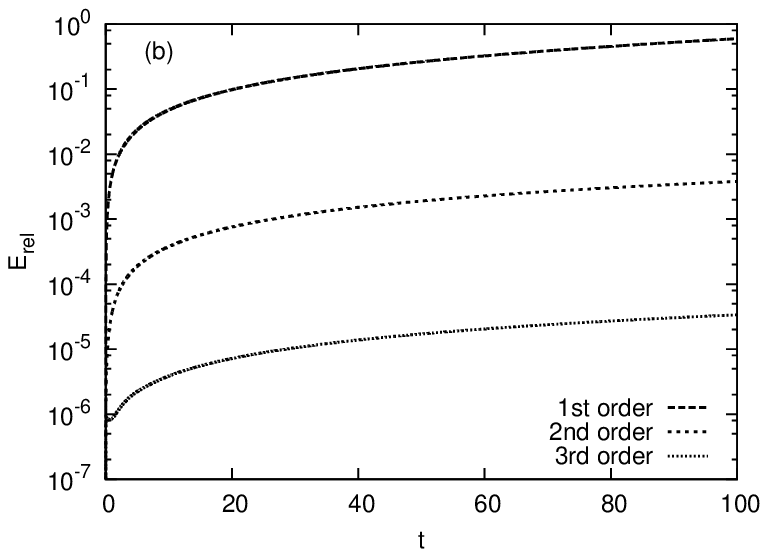}
\par\end{centering}

\caption{\label{fig:integrator-error-example}(a) The exact trajectory of a
particle starting at $\vek r_{0}=(1,0)$ with $\vek w_{0}=\vek 0$
and the parameters $R=0.75$ and $S=0.3$. The dots show the position
at integer time units. Also shown are the approximations of first,
second and third order for $h=10^{-2}$, where the latter two are
overlapped by the exact trajectory and are thus not visible. (b) The
relative error of the the numerical solutions obtained by the first,
second and third order schemes (\ref{eq:first-order-integrator})-(\ref{eq:third-order-integrator})
with $h=10^{-2}$.}
\end{figure}

To test the accuracy of the whole integration scheme, the motion of
a particle in the flow $\vek u(\vek r)=\left|\vek r\right|\vek e_{\varphi}$
(rigid body rotation) will be considered. Fortunately, in this case
there is an analytical solution for the full Maxey-Riley equation
found by Candelier et al. \citep{Angilella2004}. Qualitatively, the
solution is a spiraling motion outwards or inwards depending on whether
the density of the particle is larger or smaller then that of the
fluid, i.e. $R<1$ or $R>1$. Asymptotically the distance of the particle
from the center grows exponentially, $\left|\vek r(t)\right|\sim\exp(\lambda t)$.
The ejection rate $\lambda$ depends on the presence of the history
force and thus the trajectories of particles with memory and without
memory deviate rather quickly. This makes this flow a good choice
for a test of the integration scheme as an inaccurate computation
of the history force is expected to lead to strong deviations from
the analytically known trajectories.

Figure~\ref{fig:integrator-error-example}a shows the exact solution
together with the numerical solutions of first, second and third order
obtained by (\ref{eq:first-order-integrator})-(\ref{eq:third-order-integrator})
with $h=10^{-2}$. Only the the first order approximation is visible,
whereas the second and third order ones are overlapped by the exact
trajectory. To understand this let us examine the relative error
\[
E_{\mm{rel}}\left(t,h\right)=\frac{\left|\vek r(t)-\vek r_{\mm{num}}(t,h)\right|}{\left|\vek r(t)\right|},
\]
where $\vek r_{\mm{num}}(t,h)$ is the numerical and $\vek r(t)$
the exact solution. Figure~\ref{fig:integrator-error-example}b shows
this quantity for $h=10^{-2}$. We see that the error improves by
approximately two orders of magnitude for each additional order of
the scheme, thus explaining the overlapping of the second and third
order approximations by the exact solution in figure~\ref{fig:integrator-error-example}a.
Figure~\ref{fig:integrator-error-example}b also gives information
about the quality of the approximation as a function of time. For
example, at $t=100$ the first order approximation has a very large
error of ca. $60\%$, whereas the second and third order approximations
are rather accurate with errors of ca. $0.4\%$ and $0.003\%$.  At
$t=100$ the distance of the particle from the center is $\left|\vek r(100)\right|\approx31$
whereas for a particle without memory (i.e. when the history force
is neglected) it is $\approx\!476$, showing that the history force
has a strong influence on the particle's motion. Therefore an accurate
computation of the history force is essential for a high precision
approximation, as obtained by the second and third order schemes.

\begin{figure}
\begin{centering}
\includegraphics[height=0.3\paperwidth]{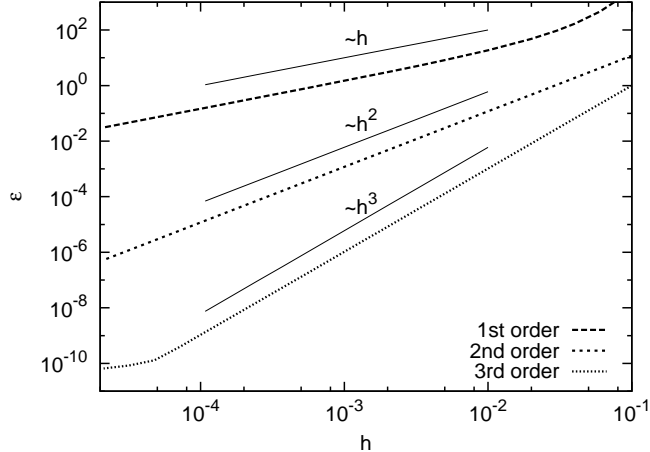}
\par\end{centering}

\caption{\label{fig:Scaling-integrator}Scaling of the global error $\varepsilon(h)$
as a function of the time-step $h$. The parameters of the particle
are $R=0.75$ and $S=0.3$; it started at $\vek r_{0}=(1,0)$ with
$\vek w_{0}=\vek 0$.}
\end{figure}

To examine the dependence of the error on the width of the time-step
let us again use the global error
\[
\varepsilon(h)=\max_{t\in[0,100]}\left|\vek r(t)-\vek r_{\mm{num}}(t,h)\right|,
\]
where $\vek r_{\mm{num}}(t,h)$ is the numerical and $\vek r(t)$
the exact solution. Figure~\ref{fig:Scaling-integrator} shows that
the error of the $m$-th order scheme scales as $h^{m}$. The global
error is expected to be proportional to the number of time-steps,
i.e. for a one-step error of $\mathcal{O}(h^{m+1})$ we expect the
global error to behave as $\, t_{\max}/h\,\mathcal{O}(h^{m+1})=\mathcal{O}(h^{m})$,
where $t_{\max}$ is the integration length. Thus the benchmark shown
in Figure~\ref{fig:Scaling-integrator} confirms the order of the
schemes.

From the dependence of $\varepsilon$ on $h$ we can see that rather
small global errors are achievable with moderately small time-steps
when the second or third order scheme is used. Let us exemplify the
importance of the higher order schemes for the computational costs
with the measurements shown in Figure~\ref{fig:Scaling-integrator}.
Suppose we would like to achieve a maximal global error of $\varepsilon=1$,
which corresponds to a relative error of approximately $\varepsilon/\left|\vek r(100)\right|\approx3\%$.
Then we would have to choose at least $h_{1}\approx8\cdot10^{-4}$,
$h_{2}\approx3\cdot10^{-2}$ and $h_{3}\approx10^{-1}$ for the first,
second and third order scheme (see Figure~\ref{fig:Scaling-integrator}).
This would lead to $N_{1}=100/h_{1}\approx10^{5}$, $N_{2}\approx3\cdot10^{3}$
and $N_{3}\approx10^{3}$ necessary time-steps for the three schemes.
The reduction in the number of time-steps is considerable. However
the reduction in the computational costs are even more dramatic as
they are proportional to $N^{2}$, e.g. using the third order scheme
would reduce the computational cost roughly by a factor of $N_{1}^{2}/N_{3}^{2}\approx10^{4}$
compared to the first order scheme. This ratio would become even higher
when we go to smaller error bounds. It should be emphasized here that
all three integration schemes have basically the same computational
cost per time-step. This is because the history force is computed
by a weighted sum in all three cases (\ref{eq:first-order-quadrature}),(\ref{eq:second-order-quadrature}),(\ref{eq:third-order-quadrature})
and the coefficients can be precomputed as discussed above.

\section{Stability of the Integration Scheme}

An important property of numerical algorithms is stability, i.e. errors
remain bounded during the iteration of the algorithm. For ordinary
differential equations numerical stability is usually determined by
applying the integration scheme to the equation 
\begin{equation}
\frac{\mathrm{d}w}{\mathrm{d}t}=-kw,\label{eq:no-memory}
\end{equation}
and verifying whether the numerical solution converges to zero. To
check the stability of our schemes we use the equation 
\begin{equation}
\frac{\mathrm{d}w}{\mathrm{d}t}=-k\left(w+\dt{}\int_{t_{0}}^{t}\frac{w(\tau)}{\sqrt{t-\tau}}\,\mm d\tau\right),\label{eq:memory}
\end{equation}
which is the Maxey-Riley equation (\ref{eq:MR-Equations-dimensionless-2})
in still fluid ($u=0$) with $R=\pi k/3$ and $S=\pi/3$. The solution
of this equation converges to zero algebraically ($\sim t^{-1.5}$)
in contrast to an exponential convergence for (\ref{eq:no-memory}).
In general $k$ is a complex number. However, here the analysis is
restricted to purely real and positive values of $k$. In this case,
$k$ can be set to $1$ by rescaling the time and we can analyze stability
as a function of the time-step $h$ only. 

Applying the integration schemes (\ref{eq:first-order-integrator})-(\ref{eq:third-order-integrator})
to the test equation (\ref{eq:memory}) yields in each case a recurrence
relation for $w_{n}$. A recurrence relation (and thus the corresponding
numerical scheme) is said to be stable when $\lim_{n\rightarrow\infty}w_{n}=0$
for every initial condition. Without the history force $w_{n}$, depends
only on a few previous values, e.g. $w_{n-1}$, $w_{n-2}$ and $w_{n-3}$
for (\ref{eq:third-order-integrator}). In this case the stability
can be checked analytically. However with the history force, $w_{n}$
depends on all previous values and we can no longer analytically determine
the stability region. Therefore we turn to a numerical stability analysis:
the scheme is iterated for $10^{6}$ time-steps and it is checked
whether $w_{n}$ converges to zero. This procedure has been carried
out for a large number of different values of the time-step $h$ and
it has been found that $w_{n}$ either converged to zero or became
infinite. These two regimes are separated by the stability threshold
$h_{\mm{th}}$, i.e. for $h<h_{\mm{th}}$ the iterated scheme converges
to zero and is thus stable and for $h>h_{\mm{th}}$ the iterated scheme
diverges and is thus unstable. Table~\ref{tab:stability-thresholds}
shows the stability thresholds for the first, second and third order
schemes (\ref{eq:first-order-integrator})-(\ref{eq:third-order-integrator}).
For comparison the row ``without memory'' contains the stability
thresholds of the schemes without the history force, i.e. normal Adams-Bashforth
schemes%
\footnote{The computed thresholds without memory are consistent with the known
stability regions of the Adams-Bashforth methods.%
}. For the first order method the inclusion of the history force increases
the stability threshold, i.e. the scheme becomes more stable. In the
case of the second and third order schemes the inverse is true; the
stability threshold is slightly lower and the schemes are less stable
when memory is included. However the influence of the history force
on the stability of the schemes seems to be rather weak as the stability
thresholds vary only by a factor of order unity. Summing up, one can
say that the integration schemes (\ref{eq:first-order-integrator})-(\ref{eq:third-order-integrator})
seem to have very similar stability properties as the corresponding
Adams-Bashforth methods for ordinary differential equations.

\begin{table}
\begin{centering}
\begin{tabular}{|c|c|c|c|}
\cline{2-4} 
\multicolumn{1}{c|}{} & \multicolumn{3}{c|}{order of the scheme }\tabularnewline
\cline{2-4} 
\multicolumn{1}{c|}{} & 1 & 2 & 3\tabularnewline
\hline 
with memory & 4.7627 & 0.9428 & 0.3886\tabularnewline
\hline 
without memory & 2.0000 & 1.0000 & 0.5455\tabularnewline
\hline 
\end{tabular}
\par\end{centering}

\caption{\label{tab:stability-thresholds}Stability thresholds $h_{\mm{th}}$
of the numerical schemes (\ref{eq:first-order-integrator})-(\ref{eq:third-order-integrator})
compared with those without the history force in (\ref{eq:memory}).}
\end{table}

\section{\label{sec:discussion_conclusion}Discussion and Conclusion}

In this paper we developed a systematic way for the derivation of
higher order numerical integration schemes for the full Maxey-Riley
equation, including the history force. Due to the singularity of the
integrand of the history force a special numerical scheme is needed.
Explicit specifications of the numerical schemes of first, second
and third order with an accuracy of $\mathcal{O}(h^{2})$, $\mathcal{O}(h^{3})$
and $\mathcal{O}(h^{4})$, respectively, have been given. Furthermore
the correctness and the order of the schemes have been verified by
comparison with known analytical solutions. 

The accuracy of the second and third order schemes represents a substantial
improvement compared with the methods available in the literature.
As discussed above the computational cost per time-step does not depend
on the order of the scheme. Thus, by using these schemes one gets
the additional accuracy or, alternatively, the reduced number of time-steps
essentially for free. 

As mentioned in the introduction, different forms of the history force
have been proposed for the case of finite particle Reynolds numbers
$Re=a\left|\vek v-\vek u\right|/\nu$. In \citep{Mei1994,Dorgan2007}
the modified history force is based on a kernel proposed by Mei, which
decays faster then the Basset kernel for large time lags. This kernel
can be expressed as follows (in dimensionless units and to be used
in (\ref{eq:MR-Equations-dimensionless})) 
\begin{equation}
K_{\mm{Mei}}=\frac{1}{\sqrt{t-\tau}}\left\{ 1+\left[\sqrt{\frac{\pi(t-\tau)^{3}}{St^{3}}}\frac{Re^{3}}{16\left(0.75+c_{2}Re\right)^{3}}\right]^{1/c_{1}}\right\} ^{-c_{1}}.\label{eq:Meis-kernel}
\end{equation}
The parameters $c_{1}$ and $c_{2}$ have been empirically determined
in \citep{Mei1994} as $c_{1}=2$, $c_{2}=0.105$ and in \citep{Dorgan2007}
as $c_{1}=2.5$, $c_{2}=0.2$. Note that in (\ref{eq:Meis-kernel})
the Basset kernel $1/\sqrt{t-\tau}$ appears as a factor; in particular
it is the only factor with a divergent behavior. Thus we can use the
methods specified in section \ref{sec:inegrator-section} to numerically
evaluate this form of the history force by pulling the second factor
in (\ref{eq:Meis-kernel}) into $f(\tau)$,
\[
f(\tau)=\left(\frac{\mathrm{d}\vek v}{\mathrm{d}\tau}-\frac{\mathrm{d}\vek u}{\mathrm{d}\tau}\right)\left\{ 1+\left[\sqrt{\frac{\pi(t-\tau)^{3}}{St^{3}}}\frac{Re^{3}}{16\left(0.75+c_{2}Re\right)^{3}}\right]^{1/c_{1}}\right\} ^{-c_{1}},
\]
and use the quadrature schemes (\ref{eq:first-order-quadrature}),
(\ref{eq:second-order-quadrature}) and (\ref{eq:third-order-quadrature}).
Note that we can not make use of the relation (\ref{eq:dt-Trick})
as the above expression for $f$ can not be explicitly formulated
as a derivative of some function. Thus the schemes (\ref{eq:first-order-integrator})-(\ref{eq:third-order-integrator})
have to be modified for the use with this kernel (essentially the
history force has to become part of $\vek G$ in section \ref{sec:inegrator-section}).

Lovalenti and Brady \citep{Lovalenti1993} derived a history force
of the form 
\begin{equation}
\int_{-\infty}^{t}\frac{1}{\left(t-\tau\right)^{3/2}}g(t,\tau,\vek v,\vek u)\,\mm d\tau=\int_{-\infty}^{t}\frac{2}{\sqrt{t-\tau}}\frac{\mm dg}{\mm d\tau}\,\mm d\tau\label{eq:brady-history-force}
\end{equation}
where $g(t,\tau,\vek v,\vek u)$ has a complicated dependence on the
particle and fluid velocity. In the limit $\tau\rightarrow t$ one
finds $g\sim t-\tau$; thus the left integral in (\ref{eq:brady-history-force})
is well defined and the identity (\ref{eq:brady-history-force}) holds.
The right integral in (\ref{eq:brady-history-force}) again contains
the Basset kernel. Furthermore the function $\frac{\mm dg}{\mm d\tau}$
has no singularities, we therefore can use the coefficients from section
\ref{sec:inegrator-section} for a numerical evaluation of the history
force, by choosing $f=2\frac{\mm dg}{\mm d\tau}$. Thus the numerical
schemes presented here can be used with the standard Basset history
force as well as with other proposed forms of the history force.

There are several reasons limiting the wide use of the history force
in simulations of inertial particles. On the one hand there is some
disagreement on the particular form of the history force in the case
of finite particle Reynolds numbers. On the other hand there are computational
problems: the high numerical costs and the absence of high accuracy
schemes. As has been shown, the later two points can be effectively
addressed with the higher order schemes developed here. I hope that
this will resolve some of the hurdles in the research on the history
force and facilitate investigations of its role in the motion of inertial
particles.

\section*{Acknowledgment}

I would like to thank Tamás Tél and Michael Wilczek for very helpful
discussions.

\appendix

\section{Details on the derivation of the quadrature scheme\label{sec:appendix-1}}

In section \ref{sec:Quadrature-Scheme} the derivation of the quadrature
scheme was presented in a simplified, not fully detailed way. This
appendix will present the technical details and give a complete, but
somewhat laborious derivation.

To interpolate $f(\tau)$ in the interval $\left[\tau_{i},\tau_{i+1}\right]$,
Lagrangian polynomial interpolation is used: 
\[
f(\tau)=\sum_{k=0}^{m}f(\theta_{ik}^{nm})L_{ik}^{nm}\left(\tau\right)+\mathcal{O}(h^{m+1})\qquad L_{ik}^{nm}(\tau)=\prod_{\substack{l=0\\
l\neq k
}
}^{m}\frac{\tau-\theta_{il}^{nm}}{\theta_{ik}^{nm}-\theta_{il}^{nm}}.
\]
Here the full dependence of the time-points $\theta_{ik}^{nm}$ on
$n$ and $m$ has been written out explicitly. In section \ref{sec:Quadrature-Scheme}
we have chosen $\theta_{ik}^{nm}=\tau_{i-\left\lfloor m/2\right\rfloor +k}$.
Thus the dependence on $m$ and $i$ is obvious, and we will see in
a moment why a dependence on $n$ is necessary. The problem with the
above definition of $\theta_{ik}^{nm}$ is that for $i<\left\lfloor m/2\right\rfloor $
we would obtain time-points outside the integration interval $\left[t_{0},t\right]$,
e.g. $\theta_{0,0}^{nm}=\tau_{-\left\lfloor m/2\right\rfloor }=t_{0}-\left\lfloor m/2\right\rfloor h$,
and thus would have to rely on values of $f(\tau)$ that are not available.
A similar problem occurs for $i>n-m+\left\lfloor m/2\right\rfloor $.
To solve this, we need a definition of $\theta_{ik}^{nm}$ that deals
with the special cases $i<\left\lfloor m/2\right\rfloor $ and $i>n-m+\left\lfloor m/2\right\rfloor $.
For this let us define the offset $o_{i}^{nm}$ as
\[
o_{i}^{nm}=\begin{cases}
0 & 0\leq i\leq\left\lfloor m/2\right\rfloor \\
i-\left\lfloor m/2\right\rfloor  & \left\lfloor m/2\right\rfloor <i<n-m+\left\lfloor m/2\right\rfloor \\
n-m & n-m+\left\lfloor m/2\right\rfloor \leq i\leq n-1
\end{cases}
\]
and $\theta_{ik}^{nm}=\tau_{o_{i}^{nm}+k}$. The offset is defined
so that it is equal to $i-\left\lfloor m/2\right\rfloor $ (corresponding
to our naive ansatz $\theta_{ik}^{nm}=\tau_{i-\left\lfloor m/2\right\rfloor +k}$)
where possible and is set to $0$ and $n-m$ where we would obtain
time-points outside the integration interval $\left[t_{0},t\right]$. 

The interpolating polynomial for $f(\tau)$ in the interval $\left[\tau_{i},\tau_{i+1}\right]$
can now be expressed as

\[
f(\tau)=\sum_{k=0}^{m}f(\tau_{o_{i}^{nm}+k})L_{ik}^{nm}\left(\tau\right)+\mathcal{O}(h^{m+1})\qquad L_{ik}^{nm}(\tau)=\prod_{\substack{l=0\\
l\neq k
}
}^{m}\frac{\tau-\tau_{o_{ik}^{nm}+l}}{\tau_{o_{ik}^{nm}+k}-\tau_{o_{ik}^{nm}+l}}
\]
and integrated to yield
\begin{equation}
\int_{t_{0}}^{t}\mathrm{d}\tau\, K\left(t-\tau\right)f(\tau)=\sum_{i=0}^{n-1}\sum_{k=0}^{m}f(\tau_{o_{i}^{nm}+k})\underbrace{\int_{\tau_{i}}^{\tau_{i+1}}\mathrm{d}\tau\, K\left(t-\tau\right)L_{ik}^{nm}(\tau)}_{\lambda_{ik}^{nm}}+E=\sum_{i=0}^{n-1}\sum_{k=0}^{m}f(\tau_{o_{i}^{nm}+k})\lambda_{ik}^{nm}+E,\label{eq:double-sum}
\end{equation}
where $E=\int_{t_{0}}^{t}K(t-\tau){\rm d}\tau\,\mathcal{O}(h^{m+1})$
is the error term.

Let us now reorder the double sum to a single sum of the type $\sum_{j}\mu_{j}^{nm}f(\tau_{n-j})$.
As already mentioned in section \ref{sec:Quadrature-Scheme}, it turns
out as beneficial to index the coefficients $\mu_{j}^{nm}$ in reversed
order, i.e. $\mu_{0}^{nm}$ and $\mu_{n}^{nm}$ correspond to $f(\tau_{n})$
and $f(\tau_{0})$ respectively. For the following calculations we
will use the theta function, which is defined here in the following
way: $\Theta$ takes logical conditions as arguments and has the value
$1$ if the condition is satisfied and $0$ otherwise, e.g. $\Theta(i<0)$
is equal to $1$ when $i<0$. The double sum in (\ref{eq:double-sum})
can be expressed as a single sum 
\[
\sum_{i=0}^{n-1}\sum_{k=0}^{m}f(\tau_{o_{i}^{nm}+k})\lambda_{ik}^{nm}=\sum_{j=0}^{n}f(\tau_{n-j})\sum_{i,k}\Theta\left(o_{i}^{nm}+k=n-j\right)\lambda_{ik}^{nm}=\sum_{j=0}^{n}\mu_{j}^{nm}f(\tau_{n-j}),
\]
with the coefficients 
\[
\mu_{j}^{nm}=\sum_{i=0}^{n-1}\sum_{k=0}^{m}\Theta\left(o_{i}^{nm}+k=n-j\right)\lambda_{ik}^{nm}.
\]
Using the definition of $o_{i}^{nm}$, the sum over $i$ can be split
into three terms (for the purpose of a compact presentation the indices
$n$ and $m$ will be omitted and the abbreviations $a=\left\lfloor m/2\right\rfloor $
and $b=m-\left\lfloor m/2\right\rfloor $ will be used):
\[
\mu_{j}=\sum_{i=0}^{a-1}\sum_{k=0}^{m}\Theta\left(k=n-j\right)\lambda_{ik}+\sum_{i=a}^{n-b}\sum_{k=0}^{m}\Theta\left(i-a+k=n-j\right)\lambda_{ik}+\sum_{i=n-b+1}^{n-1}\sum_{k=0}^{m}\Theta\left(n-m+k=n-j\right)\lambda_{ik}.
\]
The conditions in the theta functions can be used to get rid of one
summation. For example in the first term the condition $k=n-j$ is
satisfied at most for one value of $k$ and thus $k$ can be replaced
by $n-j$. However one has to keep in mind that the condition may
be not satisfiable at all (it is satisfiable when $0\leq n-j\leq m$).
Applying this kind of reasoning to the other two terms yields
\begin{multline*}
\mu_{j}=\Theta\left(0\leq n-j\leq m\right)\sum_{i=0}^{a-1}\lambda_{i,n-j}+\sum_{k=0}^{m}\Theta\left(a\leq n-j-k+a\leq n-b\right)\lambda_{n-j-k+a,k}\\
+\Theta\left(0\leq m-j\leq m\right)\sum_{i=n-b+1}^{n-1}\lambda_{i,m-j}.
\end{multline*}
In the second term the summation over $i$ (instead of $k$) has been
removed. The satisfiability condition depends on $k$ and thus has
to remain inside the sum. Simplifying the conditions we obtain
\begin{multline*}
\mu_{j}=\Theta\left(n-m\leq j\leq n\right)\sum_{i=0}^{a-1}\lambda_{i,n-j}+\sum_{k=0}^{m}\Theta\left(m-j\leq k\leq n-j\right)\lambda_{n-j-k+a,k}+\Theta\left(0\leq j\leq m\right)\sum_{i=n-b+1}^{n-1}\lambda_{i,m-j}.
\end{multline*}
The condition in the second term can be used to narrow the summation
range, and we obtain the final expression for the coefficients $\mu_{j}^{nm}$

\begin{multline*}
\mu_{j}^{nm}=\Theta\left(n-m\leq j\leq n\right)\sum_{i=0}^{a-1}\lambda_{i,n-j}^{nm}+\sum_{k=\max(0,m-j)}^{\min(m,n-j)}\lambda_{n-j-k+a,k}^{nm}+\Theta\left(0\leq j\leq m\right)\sum_{i=n-b+1}^{n-1}\lambda_{i,m-j}^{nm}.
\end{multline*}

For the case of the standard kernel (\ref{eq:standard-kernel}) the
integrals $\lambda_{ik}^{nm}$ can be computed analytically and thus
the coefficients $\mu_{j}^{nm}$. This has been done by means of the
computer algebra system Maple and the resulting expressions for the
coefficients are given in section \ref{sec:Quadrature-Scheme}.

\begin{center}
\rule[0.5ex]{0.4\columnwidth}{1pt}
\par\end{center}

\bibliographystyle{elsarticle-num}
\bibliography{integral_term}

\end{document}